\documentclass[pra,twocolumn,showpacs]{revtex4}
\usepackage{amsmath,amssymb,mathrsfs}
\usepackage{psfrag}
\usepackage{graphicx}
\usepackage{graphics}
\usepackage{epsfig}
\usepackage{bm}
\usepackage{color}
\usepackage{verbatim,color,ulem}

\newcommand{\beq}{\begin{equation}}
\newcommand{\eeq}{\end{equation}}
\newcommand{\beqa}{\begin{eqnarray}}
\newcommand{\eeqa}{\end{eqnarray}}

\begin{document}

\newpage

\title{Rabi coupled fermions in the BCS-BEC crossover}
\author{L. Dell'Anna$^{1,2}$, F. De Bettin$^{1}$, and L. Salasnich$^{1,2,3}$}
\affiliation{$^{1}$Dipartimento di Fisica e Astronomia ``Galileo Galilei'' 
and QTech Center, Universit\`a di Padova, Via Marzolo 8, 35131 Padova, Italy \\
$^{2}$Istituto Nazionale di Fisica Nucleare, Sezione di Padova, 
Via Marzolo 8, 35131 Padova, Italy\\
$^{3}$Istituto Nazionale di Ottica del Consiglio Nazionale delle Ricerche, 
via Nello Carrara 2, 50019 Sesto Fiorentino, Italy}


\begin{abstract}
We investigate the three-dimensional 
BCS-BEC crossover in the presence of  a Rabi coupling which strongly affects 
several properties of the system, 
such as the chemical potential, the pairing gap and the superfluid density. 
We determine the critical interaction strength, below which the system 
is normal also at zero temperature.  
Finally, we calculate the effect of the Rabi coupling on the critical 
temperature of the superfluid-to-normal phase transition by using 
different theoretical schemes. 
\end{abstract}

\pacs{67.85.Lm, 74.20.Fg, 47.37.+q}

\maketitle

\section{Introduction}

An extremely important achievement in the field of ultracold atoms 
has been the 
realization of the crossover from the Bardeen--Cooper--Schrieffer (BCS) 
superfluid phase of loosely bound pairs of fermions to the Bose--Einstein 
condensate (BEC) of tightly bound composite bosons~\cite{p3}. 
Recently a renewed interest in this field has been triggered by
a breakthrough experiment~\cite{p5} 
showing that the spin of an atom could be coupled to its 
center-of-mass motion by dressing two atomic spin states with a pair 
of laser beams. This technique has been then adopted in other experimental
investigations of bosonic~\cite{p6} and fermionic~\cite{p7} atomic gases with 
artificial spin-orbit 
and Rabi coupling. Triggered by this pioneering remarkable experiment 
in the last few years, a large number of theoretical 
papers have analyzed, within~a mean field approach, 
the effect of spin-orbit couplings of Rashba~\cite{p8} and Dresselhaus~\cite{p9} type, 
often with the inclusion of a Rabi term, in~
the condensates~\cite{p10,p11,p12,p13,p14,p15} 
and in the BCS--BEC crossover of superfluid fermions~\cite{p16,p17,p18,p19,
p20,p21,p22,p23,p24,p25,p26,p27,p28,p29,p30,p31,dellanna2021,powell2022}. 
In particular, a spin-orbit coupling can turn a first-order phase transition, 
driven by a Rabi coupling, into a second-order one.

The aim of this paper is, instead, the study of an ultracold gas of 
purely Rabi coupled fermionic atoms interacting via a two body contact potential. 
We consider a gas of identical atoms characterized by two 
hyperfine states. The~entire atomic sample is continuously irradiated 
by a wide laser beam. The~laser frequency is in near resonance with 
the Bohr frequency of the two hyperfine states of each atom. 
In this way, there is persistent periodic transition between the two atomic quantum states with frequency 
$\omega_R$, that is, the Rabi frequency which is proportional to the atomic dipole moment 
of the transition and to the amplitude of the laser electric field. 
The itinerant ferromagnetism of 
repulsive fermions with Rabi coupling was studied 
 in both two~\cite{penna2017} and three~\cite{penna2019} spatial dimensions. 
Here, instead, we want to investigate the interplay of Rabi coupling 
and attractive interaction for fermions in the three-dimensional 
BCS--BEC crossover. It is important to stress that 
if one considers a new spin basis of symmetric and 
anti-symmetric superpositions of the bare spin states, then the Rabi
term behaves as an effective Zeeman field that breaks the
balance of the two new spin components (see, for~instance, Ref.~\cite{lepori18}). This spin-imbalanced attractive 
fermionic system has been previously investigated~\cite{liu,marchetti}. 
The current work adds new insights into the problem, not only because 
the physical setup is different, but~also because we analyze in detail,  
as a function of the Rabi coupling, the~critical interaction strength 
below which the system is in the normal phase also at zero temperature, 
the equation of state, the~superfluid fraction, and~two 
alternative ways to determine the beyond-mean-field critical temperature. 
In Section II we introduce Rabi coupling in the model of attractive 
fermions. In~Section III we investigate the problem 
at the mean field level, also considering an improved 
determination~\cite{babaev} of the critical temperature of 
the superfluid-to-normal phase transition. In~Section IV we consider 
beyond-mean-field corrections and use them 
to calculate the critical temperature with the inclusion 
of Gaussian fluctuations~\cite{dellanna2021,nozieres}. 

\section{The model}

Our ultracold Fermi gas model is enriched with the 
addition of Rabi coupling, which enables the spin 
of the particles involved to flip. The Euclidean action, omitting the explicit 
dependence of the fermionic fields $\psi_{\sigma}({\bf r},\tau)$ 
($\sigma=\uparrow , \downarrow$) on space ${\bf r}$ and imaginary 
time $\tau$, then, reads
\beqa
S[\bar{\psi},\psi] &=& 
\int_0^\beta d\tau\int_V d^3{\bf r}\Big[\sum_\sigma\bar{\psi}_\sigma
\Big(\partial_\tau-\frac{\nabla^2}{2m}-\mu\Big)\psi_\sigma 
\nonumber 
\\
&-&g\bar{\psi}_\uparrow\bar{\psi}_\downarrow\psi_\downarrow\psi_\uparrow
+\omega_R(\bar{\psi}_\uparrow\psi_\downarrow+
\bar{\psi}_\downarrow\psi_\uparrow)\Big] ,  
\label{eq:RabiAction}
\eeqa
where $\beta=1/(k_BT)$ with $T$ the temperature and $k_B$ 
the Boltzmann constant, $\mu$ is the chemical potential and $V$ is the volume. 
Notice that in this paper we set $\hbar=1$. 
Within the path integral formalism, 
the partition function of the system is given by 
\beq 
Z = \int D[\bar{\psi},\psi] \, e^{-S[\bar{\psi},\psi]} ,   
\eeq
and from the partition function $Z$ all the thermodynamical quantities 
can be derived. 

We perform an Hubbard-Stratonovich transformation, so that the model can be 
rewritten in terms of a new action depending also on a new spinless 
complex field $\Delta(\textbf{r},\tau)$:
\beqa
S[\bar{\Delta},\Delta,\bar{\psi},\psi] &=& \int_0^\beta d\tau\int_Vd^{3}{\bf r}
\Big[\frac{|\Delta(\textbf{r},\tau)|^2}{g}
\\
&-&\frac{1}{2}\bar{\Psi}(\textbf{r},
\tau)G^{-1}\Psi(\textbf{r},\tau)\Big] 
+ \beta\sum_\textbf{p}\xi_\textbf{p},
\nonumber
\eeqa
where $\xi_{\bf p}={\bf p}^2/(2m)-\mu$ is the shifted 
single-particle energy and the Nambu spinors take the form
\beq
\Psi(\textbf{r},\tau) = \begin{pmatrix}
\psi_\uparrow(\textbf{r},\tau) \\
\bar{\psi}_\downarrow(\textbf{r},\tau) \\
\psi_\downarrow(\textbf{r},\tau) \\
\bar{\psi}_\uparrow(\textbf{r},\tau)
\end{pmatrix}
\eeq
while the inverse fermionic propagator is
\beq
\label{G-1}
G^{-1}= 
\left( 
\begin{smallmatrix}
-\partial_\tau +\frac{\nabla^2}{2m}+\mu & \Delta(\textbf{r},\tau) & 
-\omega_R & 0 \\
\bar{\Delta}(\textbf{r},\tau) & -\partial_\tau-\frac{\nabla^2}{2m}-\mu & 0 
& \omega_R \\
-\omega_R & 0 & -\partial_\tau +\frac{\nabla^2}{2m}+\mu & 
-\Delta(\textbf{r},\tau) \\
0 & \omega_R & -\bar{\Delta}(\textbf{r},\tau) & -\partial_\tau 
-\frac{\nabla^2}{2m}-\mu
\end{smallmatrix}
\right) .
\eeq

The action is now Gaussian in the fermionic degrees of freedom, 
therefore, we can integrate over them obtaining an effective theory for 
the complex field $\Delta$, whose action reads
\beqa
\label{Seff}
S_{\textrm{eff}}[\bar{\Delta},\Delta] &=& \int_0^\beta d\tau\int_V d^{3}{\bf r}
\frac{|\Delta(\textbf{r},\tau)|^2}{g} 
\\
&-& \frac{1}{2} \textrm{Tr} \ln (G^{-1}) + \beta\sum_\textbf{p}\xi_\textbf{p}. 
\nonumber 
\eeqa
It is also convenient and useful for what follows to write $G^{-1}$ in Fourier 
space, after denoting the four momenta by capital letters, like  
$P=(i\nu_n,\textbf{p})$,  introducing the fermionic Matsubara frequencies
$\nu_n = \frac{(2n+1)\pi}{\beta}$, with $n\in\mathbb{Z}$. 
From Eq.~(\ref{G-1}) we find that the Fourier components of the inverse 
fermionic propagator $G^{-1}_{KP}$ reads, therefore,
\beq
\left( 
\begin{smallmatrix}
(i\nu_n-\xi_\textbf{p})\delta_{K,P} & \Delta_{K+P} & -\omega_R
\delta_{K,-P} & 0 \\
\bar{\Delta}_{K+P} & (i\nu_n+\xi_\textbf{p})\delta_{K,P} & 0 & 
\omega_R\delta_{K,-P} \\
-\omega_R\delta_{(K,-P)} & 0 & (i\nu_n-\xi_\textbf{p})\delta_{K,P} 
& -\Delta_{K+P} \\
0 & \omega_R\delta_{K,-P} & -\bar{\Delta}_{K+P} 
& (i\nu_n+\xi_\textbf{p})\delta_{K,P}
\end{smallmatrix}
\right) . 
\label{eq:InversePropRabi}
\eeq

\section{Mean field approach}

We can now performe, from Eq.~\eqref{Seff}, the saddle 
point approximation, choosing $\Delta=\Delta_0$, namely homogeneous in 
space and time and, without lack of generality, fixing it real. 
The mean field action, then, reads 
\begin{equation}
S_{\textrm{mf}} = \beta V\frac{\Delta_0^2}{g} -\frac{1}{2} 
\textrm{Tr} \ln ({\cal G}^{-1})+ \beta\sum_\textbf{p} \xi_\textbf{p},
\label{eq:3.6}
\end{equation}
where $G_0^{-1}$ is equal to $G^{-1}$ where we take $\Delta=\Delta_0$. 
It is convenient working in Matsubara representation so that
\beq
\det (G_0^{-1}) = 
\left( 
\nu_n^2+(\omega^{-}_\textbf{p})^2
\right) 
\left(\nu_n^2+(\omega^{-}_\textbf{p})^2 \right), 
\label{eq:determinantRabi}
\eeq
with
\beqa
\omega^{+}_\textbf{p} &=& \sqrt{\xi_\textbf{p}^2+\Delta_{0}^2} 
+ \omega_R , 
\\
\omega^{-}_\textbf{p} &=& 
\sqrt{\xi_\textbf{p}^2+\Delta_{0}^2} - \omega_R . 
\label{eq:exenergiesRabi}
\eeqa
These two energies correspond to the poles of the fermionic propagator 
after a Wick rotation, meaning that they are the single particle excitation 
energies of the theory, and they differ from the case without Rabi 
coupling only by a constant shift $\omega_R$. 
The presence of Rabi coupling splits the excitation energies 
into two different energy 
levels separated by a shift $2\omega_R$. It is immediately clear that 
$\omega^-_\textbf{p}$ may take negative values, which is somewhat 
unexpected. This may happen for $\Delta_0<\omega_R$, a regime which is 
unphysical, as we will see, unless $\Delta_0 = 0$.

The mean-field grand potential is given by 
\beq
\label{Omf}
\Omega_\textrm{mf} = k_BT \, S_\textrm{mf} \; . 
\eeq
Explicitly we have 
\beqa
\Omega_\textrm{mf} = 
V\frac{\Delta_0^2}{g} -\frac{1}{2} \sum_{\nu_n,\textbf{p}} \Big[ 
\ln\big( \nu_n^2+(\omega^+_\textbf{p})^2\big) 
\nonumber \\+ 
\ln\big( \nu_n^2+(\omega^-_\textbf{p})^2 \big) \Big] , 
\eeqa
where we used $\textrm{Tr}\ln (G_0^{-1})=\ln\det (G_0^{-1})$ and 
Eq. (\ref{eq:determinantRabi}). 
After summing over the Matsubara frequencies, we get 
\beqa
{\Omega_\textrm{mf}\over V} = {\Delta_0^2\over g} - {k_BT\over 2V}
\sum_{\bf p}\Big\{
\ln\Big[2\big(1+\cosh(\beta\omega^+_\textbf{p})\big)\Big]
\nonumber 
\\
+  \ln\Big[2\big(1+\cosh(\beta\omega^-_\textbf{p})\big)\Big] \Big\} . 
\label{eq:grandpotentialRabi}
\eeqa

\subsection{Gap and number equations}

Minimizing the mean-field grand potential $\Omega_\textrm{mf}$ 
with respect to $\Delta_0$ we obtain the so-called gap equation 
\beq 
\frac{1}{g} = \frac{1}{4V}\sum_\textbf{p}\left[\frac{\tanh\big(\frac{\beta}{2}
\omega^{+}_\textbf{p}\big)}{\sqrt{\xi_\textbf{p}^2+\Delta_0^2}}
+\frac{\tanh\big(\frac{\beta}{2}\omega^{-}_\textbf{p}\big)}
{\sqrt{\xi_\textbf{p}^2+\Delta_0^2}}\right] . 
\label{eq:MFgapeqRabi}
\eeq
This equation is divergent in the ultraviolet and requires 
a regularization of the interaction strength $g$, namely 
\begin{equation}
\frac{1}{g} = -\frac{m}{4\pi a_F} + \frac{1}{V}\sum_\textbf{p}
\frac{m}{\textbf{p}^2},
\end{equation}
with $a_F$ being the physical s-wave scattering length.

\begin{figure}
\includegraphics[width=7.5cm]{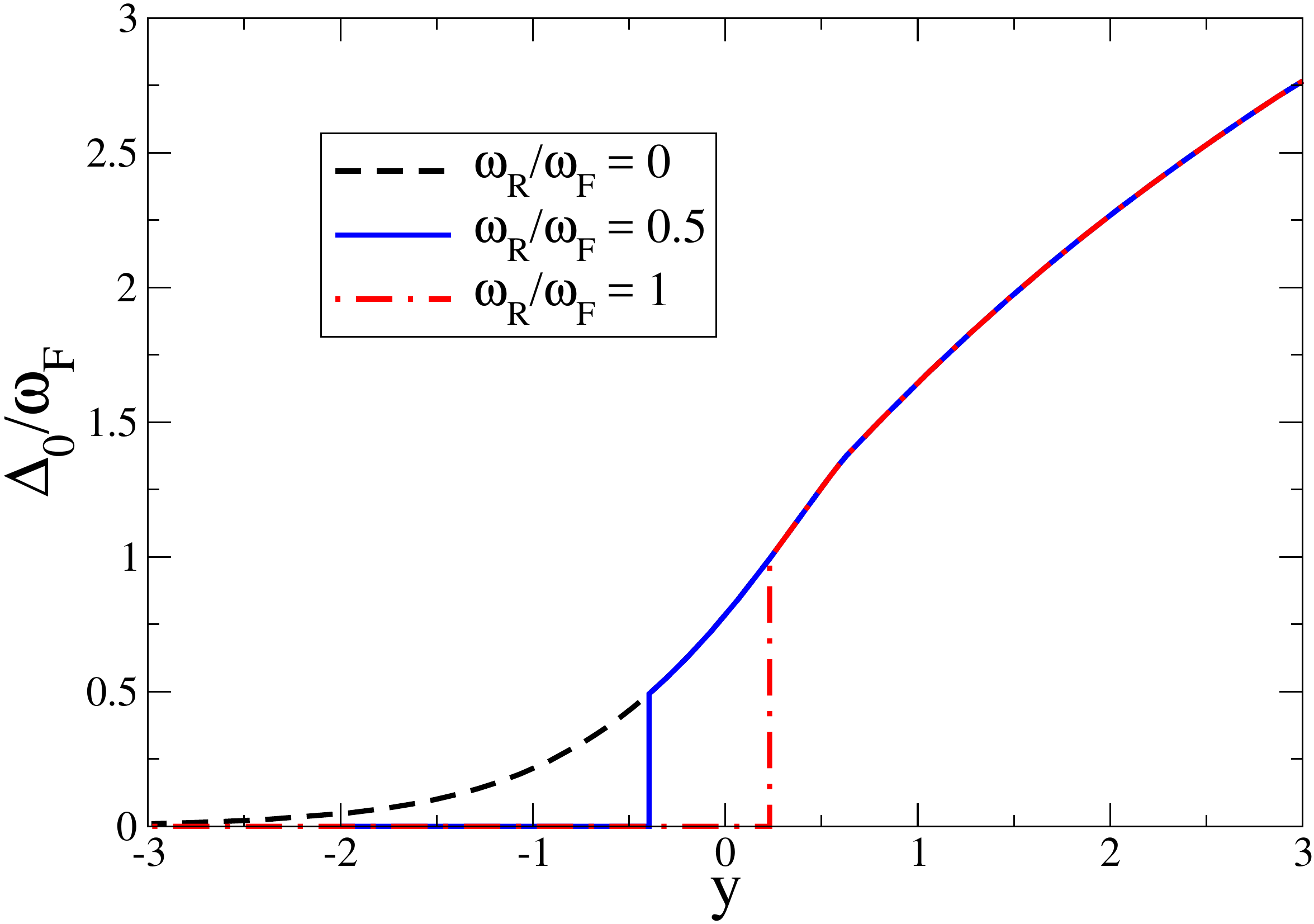}
\includegraphics[width=7.5cm]{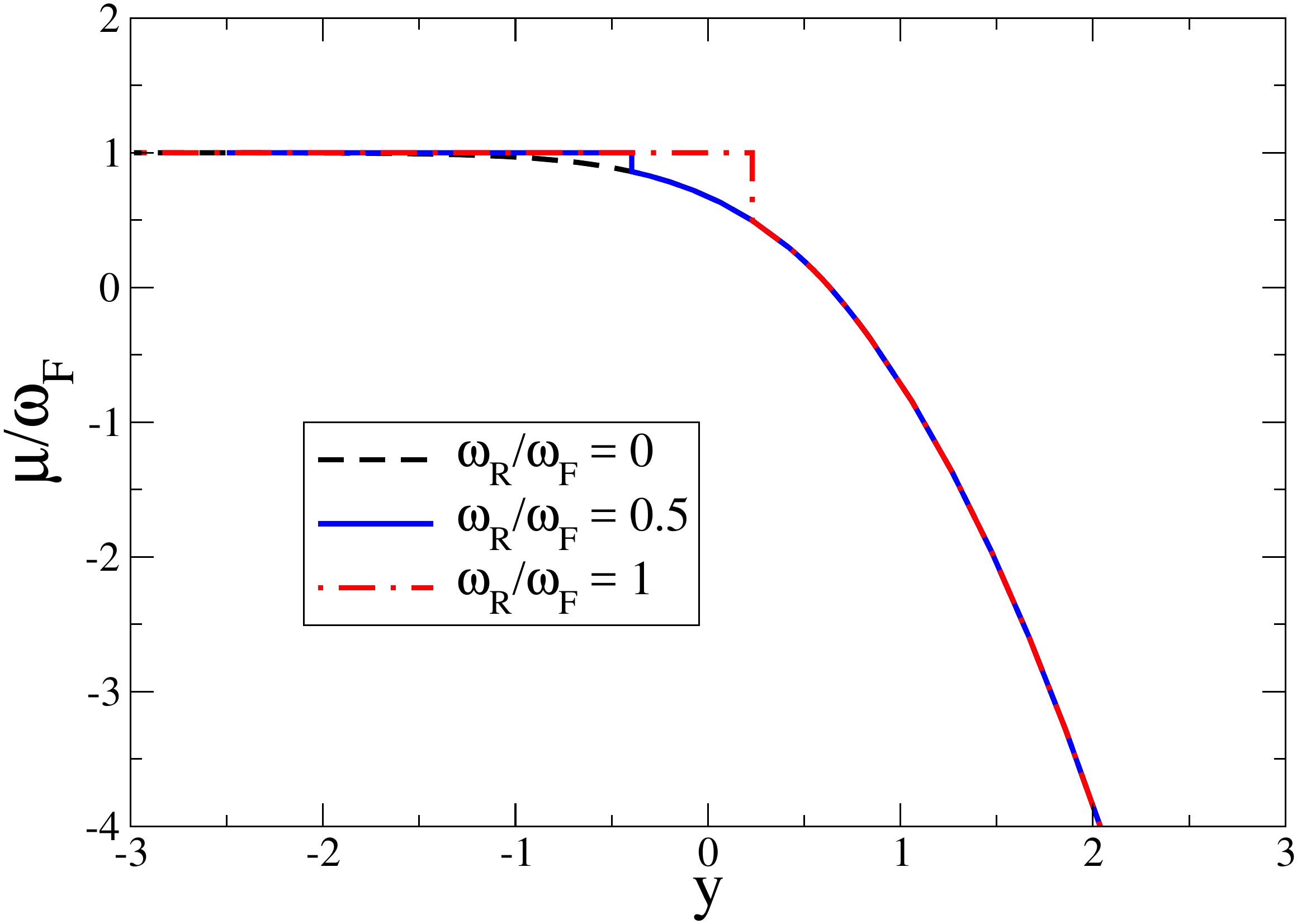}
\caption{Gap and chemical potential obtained solving Eqs.~(\ref{gap-mf}) 
and (\ref{numb-mf})  
in the in the zero-temperature limit. Upper panel: 
adimensional energy gap $\Delta_0/\omega_F$ vs 
inverse adimensional scattering length $y=1/(k_Fa_F)$. 
Lower panel: adimensional chemical potential $\mu/\omega_F$ vs  
inverse adimensional scattering length $1/(k_Fa_F)$. 
Three values of the adimensional Rabi frequency: 
$\omega_R/\omega_F = 0$ (dashed curve); 
$\omega_R/\omega_F = 0.5$ (solid curve); 
$\omega_R/\omega_F = 1$ (dotted curve). Here $k_F=(3\pi^2 n)^{1/3}$ is 
the Fermi wavenumber and $\omega_F=k_F^2/(2m)$ is the Fermi frequency.} 
\label{fig1}
\end{figure}

The total number density $n$ is instead obtained by 
\beq
n = - {1\over V}{\partial \Omega_\textrm{mf}\over \partial \mu} \; ,  
\eeq
which is the so-called number equation. 
The sum over Matsubara frequencies has the same form as the one of the 
gap equation. After some manipulations, the renormalized 
gap equation and the number equation read 
\beqa
-\frac{mV}{4\pi a_F} = \sum_\textbf{p}
\left[
\frac{\tanh\big(\frac{\beta}{2}\omega_\textbf{p}^{+}\big)
+\tanh\big(\frac{\beta}{2}\omega_\textbf{p}^{-}\big)}
{4\sqrt{\xi_\textbf{p}^2+\Delta_0^2}} 
-\frac{m}{\textbf{p}^2}
\right] 
\label{gap-mf}
\\
n V = \sum_\textbf{p}
\left[1-
\frac{\xi_\textbf{p}}{2}
\frac{\tanh\big(\frac{\beta}{2}\omega_\textbf{p}^{+}\big) 
+ \tanh\big(\frac{\beta}{2}\omega_\textbf{p}^{-}\big)}
{\sqrt{\xi_\textbf{p}^2+\Delta_0^2}}
\right] . 
\label{numb-mf}
\eeqa
The difference with respect to the case without Rabi coupling is a 
shift of $\pm\omega_R$ 
in the arguments of the hyperbolic tangents, which makes the derivation of 
analytic results more demanding.

\subsection{Zero temperature}

At zero temperature Eqs. (\ref{gap-mf}) and (\ref{numb-mf}) simplify  
because the hyperbolic tangent goes to one for  
$T\rightarrow 0^+$. However, careful attention should be paid 
studying the sign of $\omega^-_\textbf{p}$, which affects the 
form of the equations. In fact, if $\omega^-_\textbf{p}>0$ for 
any value of the momentum $\textbf{p}$, the number and gap equations 
will take the same form as the ones with no Rabi 
interaction, while if for some values of $\textbf{p}$ 
the energy $\omega^-_\textbf{p}<0$, the equations will take a 
different form, as we will show below. 
The main result of this zero-temperature 
analysis is the following: i) if $\Delta_0<\omega_R$ 
the Rabi frequency $\omega_R$ change the momenta domain of 
integration in the equations in such a way that the gap equation has 
no finite solutions; 
ii) if $\Delta_0>\omega_R$, instead, the Rabi frequency $\omega_R$
does not affect the gap and the number equations. 

In Fig. \ref{fig1} we report the plots of the energy gap $\Delta_0$ 
(upper panel) and chemical potential $\mu$ (lower panel) 
as functions of the s-wave scattering length $a_F$.  
There results are then analogous to the ones in the case without
Rabi coupling for $\Delta_0>\omega_R$, but exhibit a different behaviour 
below such a threshold. In particular, both these quantities,  
the energy gap $\Delta_0$ and the chemical potential $\mu$, for a small 
range of $1/(k_Fa_F)$, 
have two branches. However, the stable branch (associated to the 
minima of the grand potential $\Omega$) is the upper branch 
shown in the figures, superimposed to the curves of $\omega_R=0$. 
The unstable is not reported. 

\begin{figure}
\includegraphics[width=7.5cm]{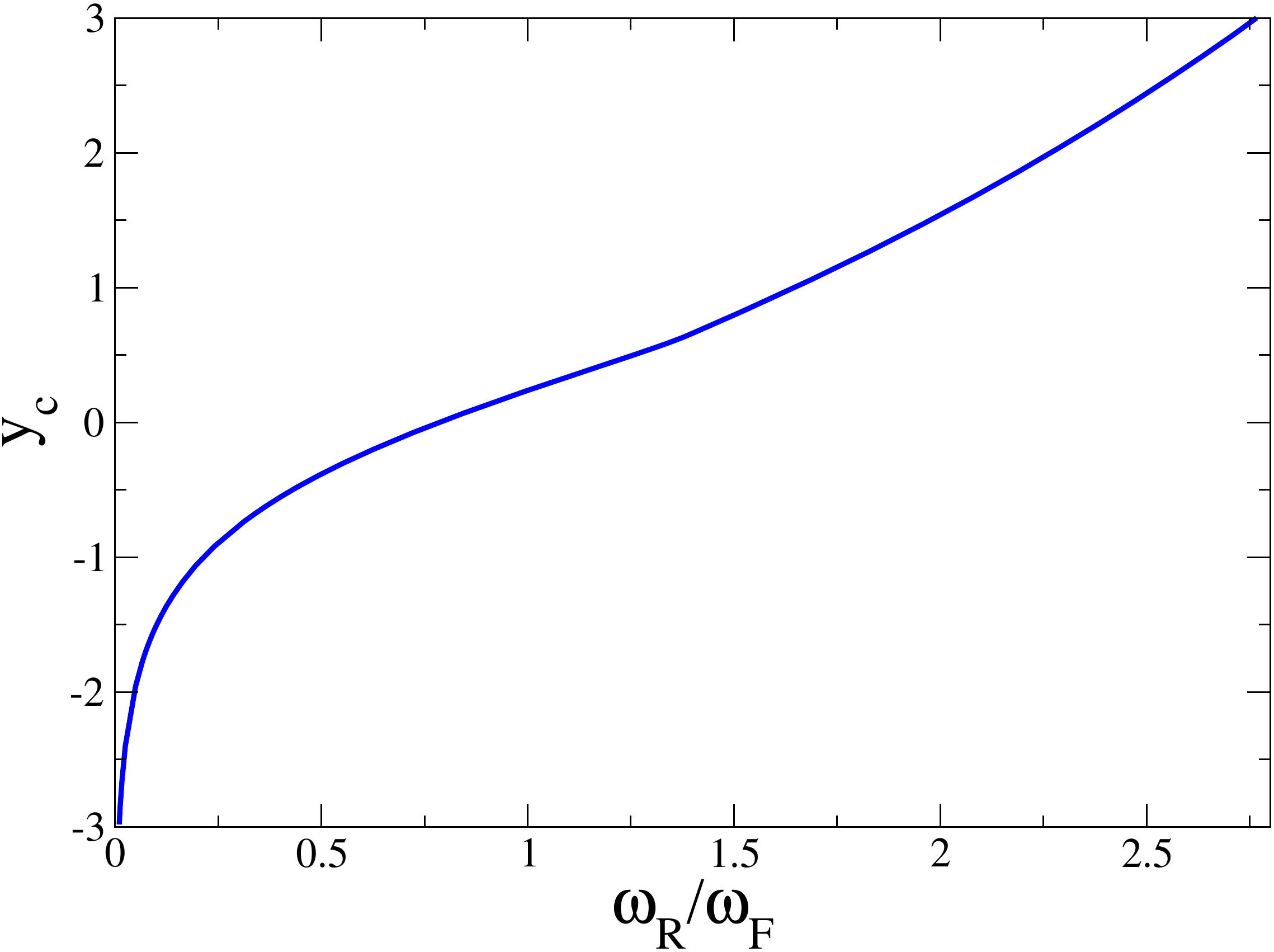}
\caption{Critical strength $y_c$ vs adimensional Rabi frequency 
$\omega_R/\omega_F$ at zero temperature. For $y\leq y_c$ the 
fermionic system is normal, i.e. the energy gap $\Delta_0=0$. 
Here $y=1/(k_Fa_F)$, where $a_F$ is the s-wave 
scattering length and  $k_F=(3\pi^2 n)^{1/3}$ is 
the Fermi wavenumber, with $\omega_F=k_F^2/(2m)$ the Fermi frequency.} 
\label{fig2}
\end{figure}

At zero temperature ($T=0$) the main effect 
of the Rabi coupling $\omega_R$ is, therefore, to make the system normal, 
i.e. with $\Delta_0=0$, for $1/(k_Fa_F)\leq y_c$, 
where $y_c=-\infty$ for $\omega_R=0$.  
The critical strength $y_c$ grows by increasing $\omega_R$. 
Specifically, given $\Delta_0(y)$ with $1/(k_F a_F)$ for $\omega_R=0$, 
$y_c$ is obtained from the condition $\Delta_0(y_c)=\omega_R$. 
This means that inverting the plot of $\Delta_0(y)$ (obtained for  
$\omega_R=0$) one gets immediately $y_c$ vs $\omega_R$, 
as shown, for the sake of completeness, in Fig. \ref{fig2}.

\subsection{Critical temperature} 

We now investigate the behaviour of the system at the critical 
temperature $T_c$, at which the energy gap $\Delta_0(T_c) = 0$. 
Let us define, for simplicity, the following adimentional quantities:
$\tilde\mu=\mu/\omega_F$,  $\tilde T_c=k_BT_c/\omega_F$, 
$\tilde\omega_R=\omega_R/\omega_F$ and $y=1/(k_Fa_F)$.
In this case the gap and the number equations can be written as follows 
\begin{eqnarray}
\label{tc1}
&& y =\frac{2 \,\tilde T_c^2}{\pi}
J_3\big(\tilde\mu,\tilde T_c,\tilde\omega_R\big) , \\
&& \tilde T_c
= \left[\frac{4}
{J_4\big(\tilde\mu,\tilde T_c,\tilde\omega_R\big)}\right]^{{2}/{3}}, 
\label{tc2}
\end{eqnarray}
where 
\beqa
J_3 &=& \int_0^{+\infty}dx \,x^2\Big[\frac{\tanh\Big(\frac{1}{2}
(x^2-{\tilde \mu\over \tilde  T_c} +{\tilde \omega_R\over \tilde T_c})
\Big)}{2(x^2-{\tilde \mu\over \tilde T_c})} 
\nonumber 
\\
&+&\frac{\tanh\Big(\frac{1}{2}(x^2-{\tilde\mu\over \tilde T_c} 
- {\tilde \omega_R\over \tilde T_c}) \Big)}
{2(x^2-  {\tilde \mu\over \tilde  T_c})}
-\frac{1}{x^2} \Big]
\label{eq:J3}
\eeqa
and
\beqa
J_4 &=& \int_0^{+\infty}dx\, x^4\Big[\frac{1}{\cosh^2\big[\frac{1}{2}
(x^2-{\tilde \mu\over \tilde  T_c} + 
{\tilde \omega_R\over \tilde  T_c})\big]} 
\nonumber 
\\
&+& \frac{1}
{\cosh^2\big[\frac{1}{2}(x^2-{\tilde \mu\over \tilde T_c} 
- {\tilde \omega_R\over \tilde  T_c})
\big]}\Big].
\label{eq:J4}
\eeqa

\begin{figure}
\includegraphics[width = 7.5cm]{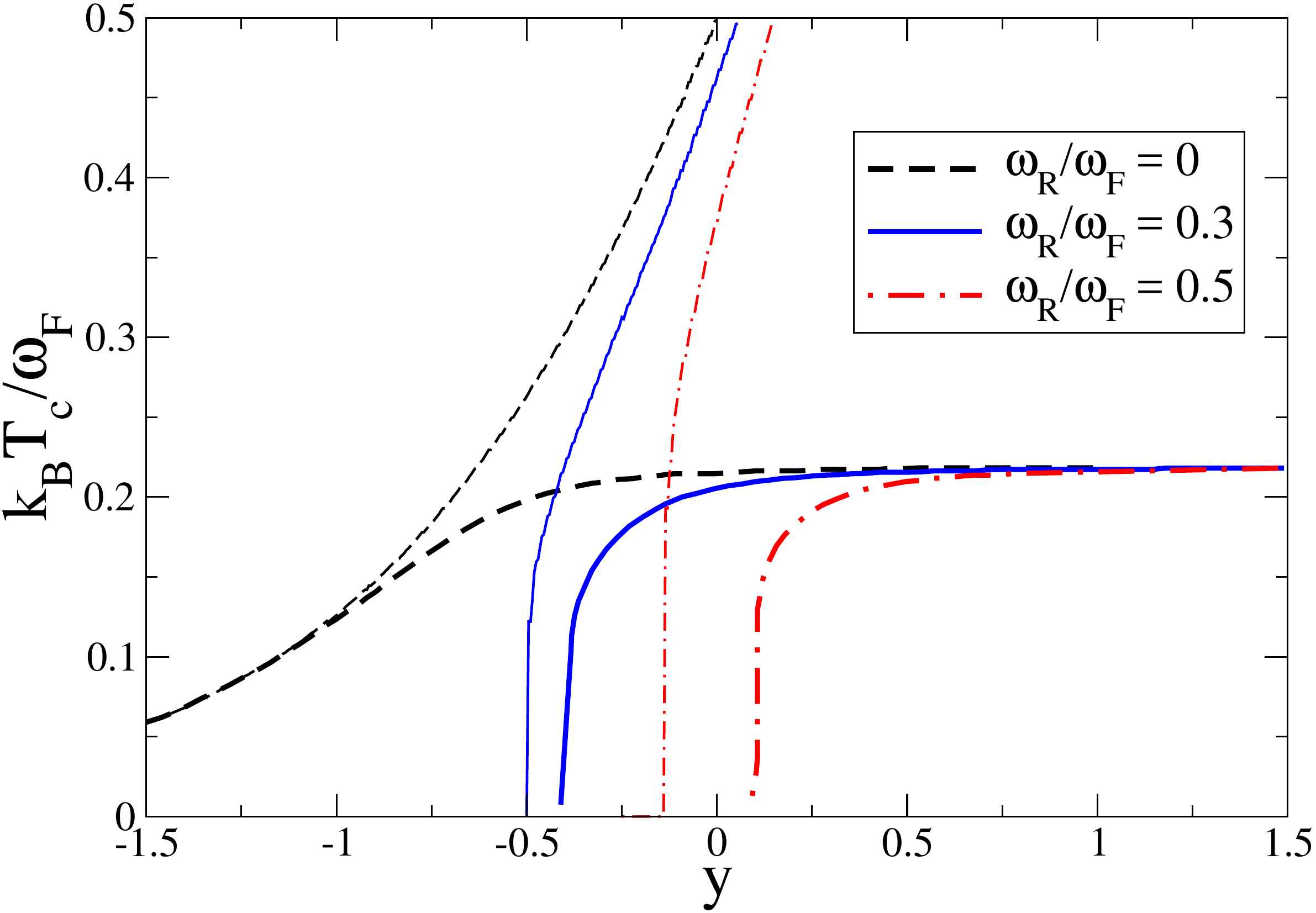}
\caption{Adimensional critical temperature $k_BT_c/\omega_F$ 
vs inverse adimensional scattering length $y=1/(k_Fa_F)$, 
with $a_F$ the s-wave scattering length, $k_F=(3\pi^2 n)^{1/3}$  
the Fermi wavenumber and $\omega_F=k_F^2/(2m)$ the Fermi frequency. 
Thin curves are the mean-field ones, 
obtained solving Eqs. (\ref{tc1}) and (\ref{tc2}),  
while thick curves are obtained from Eqs. (\ref{mah}) and (\ref{super-ns}). 
Three values of the adimensional Rabi frequency: 
$\omega_R/\omega_F = 0$ (dashed curve); 
$\omega_R/\omega_F = 0.3$ (solid curve); 
$\omega_R/\omega_F = 0.5$ (dot-dashed curve).} 
\label{fig3}
\end{figure}

Solving the coupled Eqs. (\ref{tc1}) and (\ref{tc2}) we obtain 
the critical temperature $T_c$ as a function of the inverse scattering 
length $1/a_F$ for different values of the Rabi coupling $\omega_R$. 
The results are show as thin curves in Fig. \ref{fig3}. \\
As expected, the Rabi coupling inhibits the formation of Cooper pairs: 
the stronger the Rabi coupling,  
the higher is the threshold of the 
scattering rate above which superfluidity can occur at the mean field level. 
In the strong coupling limit, instead, even in the absence of Rabi coupling, the mean field approach is expected to fail since it cannot describe the emergence of bosonic molecules which undergo condensation below a finite critical temperature. We have, therefore, to go beyond the mean field approximation.

\section{Beyond mean field} 
We are now presenting a couple of techniques which allow us to go beyond 
the mean field analysis. 
The first approach is a method based on the determination of the superfluid 
density, while the second one is based on the inclusion 
of the Gaussian fluctuations. In the latter case we will show some explicit 
results in the so-called bosonic approximation.

\subsection{By superfluid density} 
An improved determination of the critical temperature $T_c$ can be 
obtained with the method proposed by Babaev and Kleinert \cite{babaev}, 
which is the three-dimensional analog of the Nelson-Kosterlitz criterion. 
In particular, 
\beq 
k_B T_c = \alpha{n_s(T_c)\over 4m} \left({2\over n}\right)^{1/3} \; , 
\label{mah}
\eeq
where $n_s(T)$ is the mean-field superfluid density and $n$ is the 
total fermionic number density. 
Actually $J=n_s/(4m)$ is the stiffness in an effective XY model, 
$H=\frac{J}{2}\int d\textbf{r} \nabla\theta(\textbf{r})$, where $\theta$ 
is the local phase of the pairing $\Delta$.

The constant $\alpha$ is fixed to the value 
$\alpha = 2\pi/\zeta(3/2)^{2/3}$ such that $T_c$ turns out to be the exact 
value $k_B T_c =\pi/m(n/(2\zeta(3/2)))^{2/3}$ 
for $n/2$ non-interacting bosons with mass $2m$ in the deep BEC regime. 
The superfluid density $n_s(T)$ can be calculated,  
following the Landau's approach \cite{landau}, getting
\beq
n_s(T) = n + \frac{1}{6}\int\frac{d^3p}{(2\pi)^3}\frac{p^2}{m}
\Big[\frac{df_B(\omega^+_\textbf{p})}{d\omega^+_\textbf{p}}+
\frac{df_B(\omega^-_\textbf{p})} {d\omega^-_\textbf{p}}\Big] 
\label{super-ns}
\eeq
where 
\beq 
f_B(\omega) = {1\over e^{\beta\omega}-1}
\eeq
is the Bose-Einstein distribution and $n$ is the total number density. 
The thick curves of Fig. \ref{fig3} are obtained using 
Eq. (\ref{mah}) with Eq. (\ref{super-ns}), where $\Delta_0$ 
and $\mu$ are numerically determined from 
Eqs. (\ref{gap-mf}) and (\ref{numb-mf}). \\
As done previously, it is convenient to introduce adimensional  quantities, 
$\tilde\mu=\mu/\omega_F$,  $\tilde T_c=k_BT_c/\omega_F$, 
$\tilde\omega_R=\omega_R/\omega_F$ and $y=1/(k_Fa_F)$
together with $\tilde\Delta_0=\Delta_0/\omega_F$, where we recall that 
$\omega_F=k_F^2/(2m)=(3\pi^2 n)^{2/3}/(2m)$.
We can, therefore, rewrite Eq.~(\ref{mah}) as 
\beq
\tilde T_c=\frac{2}{(6\sqrt{\pi}\, \zeta(3/2))^{2/3}}\frac{n_s(\tilde T_c)}{n}
\eeq
which, in the deep BEC where all the fermions contribute to the superfluid 
density, gives $\tilde T_c\approx 0.218$, and where
\begin{eqnarray}
\hspace{-0.cm}
\nonumber 
\frac{n_s(\tilde T_c)}{n}&=& 
1 - \frac{1}{2\tilde T_c}
\int_0^{+\infty}\hspace{-0.5cm} dx\, x^4
\\&&
\hspace{-0.8cm}\left[\sum_{s=\pm 1}
\frac{e^{\frac{1}{\tilde T_c}
\Big(\sqrt{(x^2-\tilde\mu)^2+{\tilde\Delta_0^2}}
+s\,\tilde\omega_R\Big)}}
{\Big(e^{\frac{1}{\tilde T_c}
\Big(\sqrt{(x^2-\tilde\mu)^2+\tilde\Delta_0^2}
+s\,\tilde\omega_R\Big)}-1\Big)^2}
\right].
\end{eqnarray}

\subsection{Gaussian fluctuations} 

We now introduce Gaussian fluctuations in the partition function 
of the system adopting the Nozieres-Schmitt-Rink approach \cite{nozieres}. 
The aim is to derive a more precise form 
for the number equation in order to understand 
the role of quantum fluctuations on the relation between the 
chemical potential $\mu$ and the density $n$ of fermions. 

In order to introduce fluctuations we separate the 
field $\Delta(\textbf{r},\tau)$ in its homogeneous part $\Delta_0$,  
obtained minimizing the grand potential at the mean field level, and 
its small fluctuations $\eta(\textbf{r},\tau)$ around the saddle point solution 
\beq
\Delta(\textbf{r},\tau) = \Delta_0 + \eta(\textbf{r},\tau).
\eeq
After some calculations, at the leading order in 
$\eta$, the effective action Eq.~(\ref{Seff}) at the Gaussian level reads
\beq
S_{\textrm{eff}} \simeq  S_\textrm{mf} + \frac{1}{2}\sum_Q\begin{pmatrix}
\bar{\eta}_Q & \eta_{-Q}
\end{pmatrix}M_Q\begin{pmatrix}
\eta_Q \\
\bar{\eta}_{-Q}
\end{pmatrix} 
\eeq
with
\beq
M_Q= \frac{1}{g}\mathbb{I}+\chi_Q , 
\label{eq:RabiGaussMKDEF}
\eeq
where $\chi_Q$ is the contribution coming from the expansion of 
$\textrm{Tr}\ln(G^{-1})$ and $\mathbb{I}$ denotes the $2\times 2$ 
identity matrix in Nambu space. 
The components of $\chi_Q$, shown in Appendix, depend on the the quadrivector 
$Q=(i \upsilon_n,\textbf{q})$ 
where $\upsilon_n$ are bosonic Matsubara frequencies, 
$\upsilon_n = \frac{2\pi n}{\beta}$, with  $n\in\mathbb{Z}$. 

Our objective is to find an expression for the grand canonical potential 
from which we can recover a treatable expression for the contribution 
of the Gaussian fluctuations to the number equation. The effective theory 
we obtained is Gaussian, 
meaning that it can be integrated explicitly, getting the following 
partition function 
\begin{equation}
Z = e^{-S_\textrm{mf}}\prod_Q\Big[\det(M_Q)\Big]^{-1} .  
\label{eq:RabiGaussPartFunc}
\end{equation}
The grand potential, then, reads 
\beq
\Omega=\Omega_\textrm{mf} + \Omega_\textrm{g} , 
\eeq
where $\Omega_\textrm{mf}$ is given by Eq.~(\ref{Omf}) and
\beq 
\label{Og}
\Omega_\textrm{g} = k_BT\sum_Q\ln[\det(M_Q)] . 
\eeq
To compute the sum over Matsubara frequencies in 
\eqref{Og} one may analytically continue 
the argument of the sum by setting $i\upsilon_{n}\rightarrow \omega$
and transforming the sum into an integral. In this way we eventually find 
\begin{equation}
\Omega_\textrm{g} =-\frac{1}{\pi}
\sum_\textbf{q} \int_{-\infty}^{+\infty} d\omega \, 
f_B(\omega) \, {\tilde \delta}(\omega,{\bf q}) 
\end{equation}
with $f_B(\omega)$ the Bose-Einstein distribution and 
\beq 
{\tilde \delta}(\omega,{\bf q}) = 
\mbox{arctan}\left[{Im[\mbox{det}[M_{\omega,{\bf q}}]
\over Re[\mbox{det}[M_{\omega,{\bf q}}]}\right]
\eeq
the phase of the complex metrix element $M_{\omega,{\bf q}}$ 
derived from (\ref{eq:RabiGaussMKDEF}). Thus, the Gaussian 
correction to the number density reads 
\beq 
\label{ng}
n_\textrm{g} = - {1\over V} {\partial \Omega_\textrm{g}\over \partial \mu} = 
\frac{1}{\pi V}
\sum_\textbf{q} \int_{-\infty}^{+\infty} d\omega \, 
f_B(\omega) \, {\partial {\tilde \delta}
(\omega,{\bf q})\over \partial \mu} \; . 
\eeq

\begin{figure}
\includegraphics[width = 7.5cm]{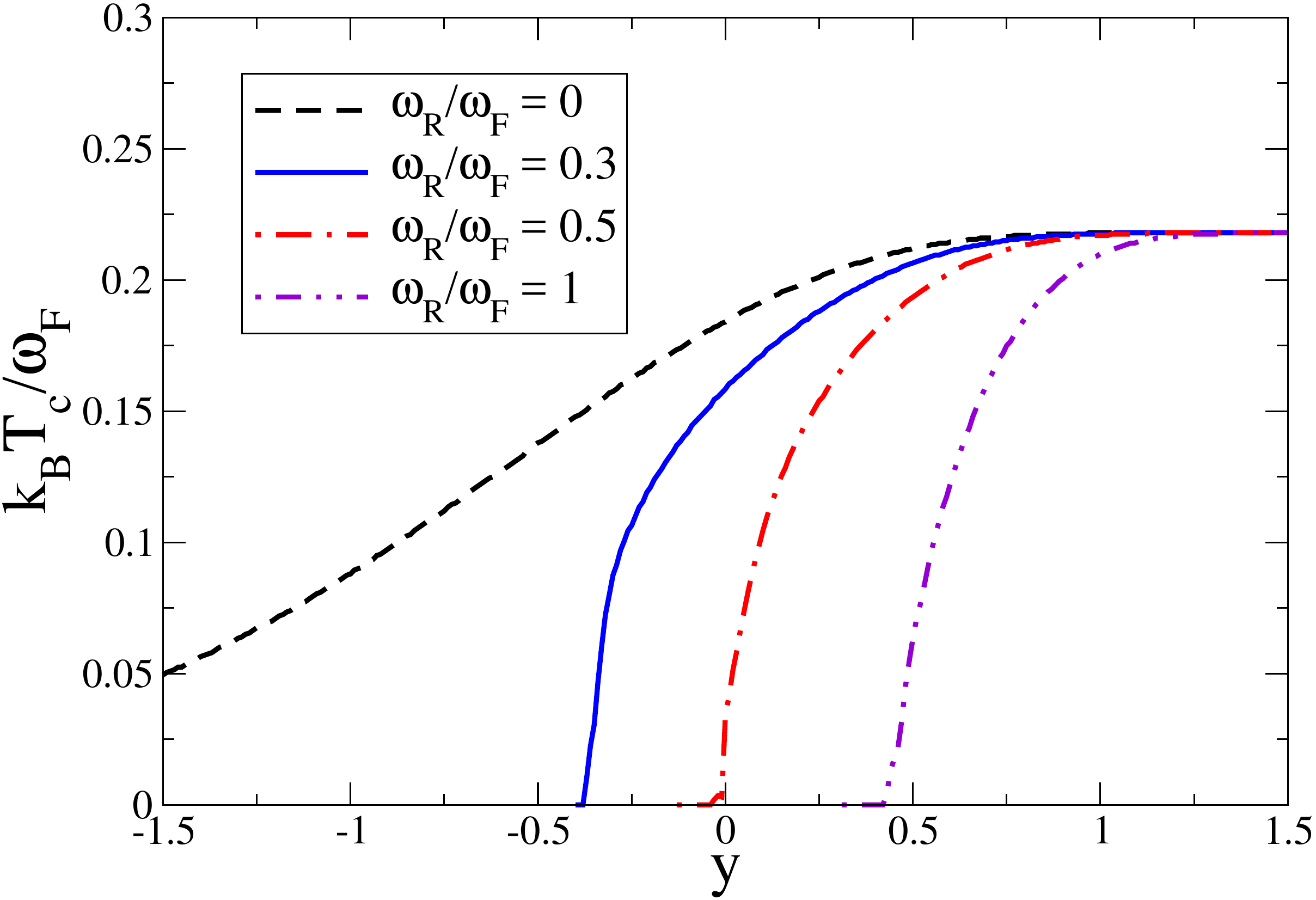}
\caption{Adimensional critical temperature $k_BT_c/\omega_F$ 
vs inverse adimensional scattering length $y=1/(k_Fa_F)$ 
by including Gaussian fluctuations within the 
bosonic approximation, i.e. solving Eqs. (\ref{gap-mf}) 
and (\ref{numb-bos}). Also here 
$a_F$ is the s-wave scattering length, $k_F=(3\pi^2 n)^{1/3}$ is 
the Fermi wavenumber and $\omega_F=k_F^2/(2m)$ is the Fermi frequency. 
Four values of the adimensional Rabi frequency: 
$\omega_R/\omega_F = 0$ (dashed curve); 
$\omega_R/\omega_F = 0.3$ (solid curve); 
$\omega_R/\omega_F = 0.5$ (dot-dashed curve); $\omega_R/\omega_F = 1$ 
(dot-dot-dashed curve).} 
\label{fig4}
\end{figure}

In the strong coupling limit $a_F\to 0^+$ 
the system becomes a gas of free bosonic dimers, made of two 
fermions with opposite spin and binding energy $E_B=-1/(ma_F^2)$. 
These bosons have mass $m_B=2$ and chemical potential $\mu_B=2\mu-E_B$. 
Indeed, as shown in Ref.~\cite{dellanna2021}, in this regime 
$\mu_B\rightarrow 0$ and 
\beq
{\partial {\tilde \delta}(\omega,{\bf q}) \over \partial \mu} 
\rightarrow \pi \delta\big(\omega - {q^2\over 2m_B} + \mu_B \big) 
\eeq
where $\delta(x)$ is the Dirac delta function. It follows 
immediately that, in the strong coupling approximation, 
also called bosonic approximation \cite{dellanna2021}, 
the number equation with Gaussian fluctuations becomes 
\beqa 
n V &=& \sum_\textbf{p} \Big[1- \frac{\xi_\textbf{p}}{2}
\frac{\tanh\big(\frac{\beta}{2}\omega_\textbf{p}^{+}\big) 
+ \tanh\big(\frac{\beta}{2}\omega_\textbf{p}^{-}\big)}
{\sqrt{\xi_\textbf{p}^2+\Delta_0^2}}\Big] 
\nonumber 
\\
&+&\sum_\textbf{q} {1\over e^{\beta({q^2\over 4m} - \mu_B)} 
- 1} \; . 
\label{numb-bos}
\eeqa
In Fig. \ref{fig4} we report the critical temperature $T_c$ 
as a fuction of the inverse scattering length $1/a_F$ obtained 
by solving Eqs. (\ref{gap-mf}) and (\ref{numb-bos}),  
setting $\mu_B=0$ and $\Delta_0=0$. 
The full calculation of Eq. (\ref{ng}) is more computationally demanding and is expected to 
deviate from the bosonic approximation only in the intermediate regime near unitarity limit ($y = 0$) 
producing a little hump in the $T_c$ profile, which is, however, a debated feature within other theoretical schemes \cite{pieri}.

\section{Conclusions}

The BCS-BEC crossover has been studied in the presence of Rabi coupling 
by using the finite-temperature path integral formalism. 
The behavior of many physical quantities 
has been studied along the whole crossover, including 
the mean-field chemical potential and energy gap at zero temperature, 
and the critical temperature at and beyond mean-field level. 
We have found that only in the deep BEC regime 
the physical properties of the system are not affected by the 
Rabi coupling. In general, also at zero temperature it exists 
a critical interaction strength, below which the system is normal. 
We have determined this critical strength as a function of the Rabi coupling. 
In the last part of the paper we have calculated the critical temperature 
of the superfluid-to-normal phase transition for different values of the 
Rabi coupling. The treatment beyond mean-field level has been carried out 
following two different procedures: the Babaev-Kleinert \cite{babaev} 
and the Nozieres-Schmitt-Rink \cite{nozieres} approaches: the 
first is based on the determination of the mean-field superfluid density as a function of the temperature while 
the second is more rigorous 
but computationally demanding. Indeed, the Nozieres-Schmitt-Rink scheme has  
been used but adopting the bosonic pair approximation \cite{dellanna2021},  
which makes the scheme more feasible numerically. 

\paragraph*{Acknowledgements.}
LS acknowledges the BIRD project "Ultracold atoms in curved geometries" 
of the University of Padova for partial support.

\section*{Appendix}
Let us write the fermionic propagator, including quantum fluctuations, 
as follows
\beq
G^{-1}=G^{-1}_{0}+\hat \eta=G^{-1}_{0}(1+G_0\hat \eta)
\eeq
where we introduce the matrix
\beq
\hat \eta_{Q}=\left( 
\begin{matrix}
0& \eta_{Q} & 0 & 0 \\
\bar{\eta}_{Q} & 0 & 0 & 0\\
0& 0 & 0
& -\eta_{Q} \\
0 & 0 & -\bar{\eta}_{Q} 
& 0
\end{matrix}
\right) . 
\label{hateta}
\eeq
so that, in the action, the $\textrm{Tr}\ln (G^{-1})$ can be written in 
the following way
\beq
\frac{1}{2}\textrm{Tr}\ln(G^{-1})=\frac{1}{2}\textrm{Tr}
\ln(G_0^{-1})+\frac{1}{2}\textrm{Tr}\ln(1+G_0\hat \eta).
\eeq
We can now expand the last term, getting the following leading term
\beq
\frac{1}{2}\textrm{Tr}\left(G_0\hat\eta G_0\hat\eta\right)=\frac{k_BT}{V}
\sum_{Q,P}\textrm{tr}(G_{0P}\hat\eta_QG_{0P+Q}\hat\eta_{-Q})
\eeq
which can be recasted, performing the matrix products, as follows
\beq
\frac{1}{2}\textrm{Tr}\left(G_0\hat\eta G_0\hat\eta\right)= 
\frac{1}{2}\sum_Q\begin{pmatrix}
\bar{\eta}_Q & \eta_{-Q}
\end{pmatrix}\chi_Q\begin{pmatrix}
\eta_Q \\
\bar{\eta}_{-Q}
\end{pmatrix} 
\eeq
where $\chi_Q$ is a $2\times 2$ matrix, introduced 
in the main text in Eq. (\ref{eq:RabiGaussMKDEF}), which, after performing 
the inverse of $G_0^{-1}$ and making the matrix products, 
can be written explicitly. This matrix is composed by the following diagonal 
terms (see Ref. \cite{tesi} for more details)
\begin{widetext}
\begin{equation}
\begin{split}
(\chi_Q)_{11} = (\chi_{-Q})_{22}=
\frac{k_BT}{V}\sum_{\textbf{p},\nu_n}\frac{[(i\nu_n-\xi_\textbf{p})((i\nu_n)^2
-\xi_\textbf{p}^2-\Delta_0^2)-\omega_R^2(i\nu_n+\xi_\textbf{p})]}
{(\nu_n^2+(\omega^+_\textbf{p})^2)(\nu_n^2
+(\omega^-_\textbf{p})^2)} \\
\times\frac{[(i\upsilon_m+i\nu_n+\xi_{\textbf{q}+\textbf{p}})
((i\upsilon_m+i\nu_n)^2-\xi_{\textbf{q}+\textbf{p}}^2-\Delta_0^2)
-\omega_R^2(i\upsilon_m+i\nu_n-\xi_{\textbf{q}+\textbf{p}})]}
{((\upsilon_m+\nu_n)^2+(\omega^+_\textbf{q+p})^2)
((\upsilon_m+\nu_n)^2+(\omega^-_\textbf{q+p})^2}  \\
+\omega_R^2\frac{k_BT}{V}
\sum_{\textbf{p},\nu_n}
\frac{[\Delta_0^2-\omega_R^2
+(i\nu_n-\xi_\textbf{p})^2]}{(\nu_n^2+(\omega^+_\textbf{p})^2)
(\nu_n^2+(\omega^-_\textbf{p})^2)} 
\frac{[\Delta_0^2-\omega_R^2+(i\upsilon_m+i\nu_n+
\xi_{\textbf{p}+\textbf{q}})^2]}{((\upsilon_m+\nu_n)^2+
(\omega^+_\textbf{q+p})^2)((\upsilon_m+\nu_n)^2
+(\omega^-_\textbf{q+p})^2)}
\end{split} 
\label{eq:RabiGaussChi12_diag}
\end{equation}
and off-diagonal terms
\begin{equation}
\begin{split}
(\chi_Q)_{12}=(\chi_Q)_{21}=\Delta_0^2\frac{k_BT}{V}\sum_{\textbf{p},\nu_n}
\frac{\big[(\nu_n^2+\xi_\textbf{p}^2)-\omega_R^2+\Delta_0^2\big]}
{(\nu_n^2+(\omega^+_\textbf{p})^2)(\nu_n^2
+(\omega^-_\textbf{p})^2)} 
\frac{\big[((\upsilon_m+\Omega_{n_F})^2+\xi_{\textbf{p}+\textbf{q}}^2)
-\omega_R^2+\Delta_0^2\big]}{((\upsilon_m+\nu_n)^2+(\omega^+_\textbf{q+p})^2)
((\upsilon_m+\nu_n)^2+(\omega^-_\textbf{q+p})^2)}  \\
-4\omega_R^2\Delta_0^2\frac{k_BT}{V}
\sum_{\textbf{p},\nu_n}
\frac{\nu_n}{(\nu_n^2+(\omega^+_\textbf{p})^2)(\nu_n^2
+(\omega^-_\textbf{p})^2)} 
\frac{(\upsilon_m+\nu_n)}{((\upsilon_m+\nu_n)^2
+(\omega^+_\textbf{q+p})^2)((\upsilon_m+\nu_n)^2
+(\omega^-_\textbf{q+p})^2)}.
\end{split} 
\label{eq:RabiGaussChi12_off}
\end{equation}
\end{widetext}

\end{document}